\documentclass[11pt]{article}
\usepackage{epsfig}
\def\nn{\nonumber}
\def\be{\begin{equation}}
\def\ee{\end{equation}}
\newcommand{\bea}{\begin{eqnarray}}
\newcommand{\eea}{\end{eqnarray}}
\newcommand{\bdm}{\begin{displaymath}}
\newcommand{\edm}{\end{displaymath}}
\long\def\symbolfootnote[#1]#2{\begingroup%
\def\thefootnote{\fnsymbol{footnote}}\footnote[#1]{#2}\endgroup}

\def\sdbeta{s_{2\beta}}

\def\sq2{\sqrt{2}}
\def\drbar{\overline{\rm DR}}
\def\msbar{\overline{\rm MS}}

\def\tb{\tan\beta}

\def\gl{\tilde{g}}
\def\mg{m_{\gl}}

\def\hsm{H_{\scriptscriptstyle{\rm SM}}}

\def\x1g{x_{1}}
\def\MS{M}

\newcommand{\smallz}{{\scriptscriptstyle Z}} 
\newcommand{\smallw}{{\scriptscriptstyle W}} %
\newcommand{\smallH}{{\scriptscriptstyle H}} %
\newcommand{\smallr}{{\scriptscriptstyle R}} %
\newcommand{\smalla}{{\scriptscriptstyle A}} %
\newcommand{\mz}{m_\smallz}
\newcommand{\mw}{m_\smallw}
\newcommand{\mH}{m_\smallH}
\newcommand{\ma}{m_\smalla}

\newcommand{\muR}{\mu_\smallr}
\newcommand{\HHu}{{\cal H}_\smallH^{1\ell}}
\newcommand{\HHd}{{\cal H}_\smallH^{2\ell}}
\newcommand{\Hhu}{{\cal H}_h^{1\ell}}
\newcommand{\Hhd}{{\cal H}_h^{2\ell}}

\def\mt{m_t}
\def\stu{\tilde{t}_1}
\def\std{\tilde{t}_2}

\def\tu{m_{\tilde{t}_1}^2}
\def\td{m_{\tilde{t}_2}^2}
\def\ti{m_{\tilde{t}_i}^2}
\def\tuq{m_{\tilde{t}_1}^4}
\def\tdq{m_{\tilde{t}_2}^4}

\def\sdt{s_{2\theta_t}}
\def\cdt{c_{2\theta_t}}

\def\mb{m_b}

\def\sdb{s_{2\theta_b}}

\begin{document}

\begin{titlepage}


{\flushright{
        \begin{minipage}{5cm}
          RM3-TH/12-5
        \end{minipage}        }

}
\renewcommand{\thefootnote}{\fnsymbol{footnote}}
\vskip 2cm
\begin{center}
\boldmath
{\LARGE\bf On the NLO QCD corrections to the production \\[7pt]
of the heaviest neutral Higgs scalar in the MSSM}\unboldmath
\vskip 1.cm
{\Large{G.~Degrassi$^{a}$, S.~Di~Vita$^{a}$ and P.~Slavich$^{b}$}}
\vspace*{8mm} \\
{\sl ${}^a$
    Dipartimento di Fisica, Universit\`a di Roma Tre and  INFN, Sezione di
    Roma Tre \\
    Via della Vasca Navale~84, I-00146 Rome, Italy}
\vspace*{2.5mm}\\
{\sl ${}^b$  LPTHE, 4, Place Jussieu, F-75252 Paris,  France}
\end{center}
\symbolfootnote[0]{{\tt e-mail:}}
\symbolfootnote[0]{{\tt degrassi@fis.uniroma3.it}}
\symbolfootnote[0]{{\tt divita@fis.uniroma3.it}}
\symbolfootnote[0]{{\tt slavich@lpthe.jussieu.fr}}

\vskip 0.7cm

\begin{abstract}
  We present a calculation of the two-loop top-stop-gluino
  contributions to Higgs production via gluon fusion in the MSSM. By
  means of an asymptotic expansion in the heavy particle masses, we
  obtain explicit and compact analytic formulae that are valid when
  the Higgs and the top quark are lighter than stops and gluino,
  without assuming a specific hierarchy between the Higgs mass and the
  top mass. Being applicable to the heaviest Higgs scalar in a
  significant region of the MSSM parameter space, our results
  complement earlier ones obtained with a Taylor expansion in the
  Higgs mass, and can be easily implemented in computer codes to
  provide an efficient and accurate determination of the Higgs
  production cross section.

\end{abstract}
\vfill
\end{titlepage}    
\setcounter{footnote}{0}


\section{Introduction}

With the coming into operation of the Large Hadron Collider (LHC), a
new era has begun in the search for the Higgs boson(s). This search
requires an accurate control of all the Higgs production and decay
mechanisms, including the effects due to radiative corrections
\cite{yellowbooks}. At the LHC the main production mechanism for the
Standard Model (SM) Higgs boson, $\hsm$, is the loop-induced gluon
fusion mechanism \cite{H2gQCD0}, $gg \to \hsm$, where the coupling of
the gluons to the Higgs is mediated by loops of colored fermions,
primarily the top quark. The knowledge of this process in the SM
includes the full next-to-leading order (NLO) QCD corrections
\cite{H2gQCD1,SDGZ,HK,babis1,ABDV,BDV}; the next-to-next-to-leading
order (NNLO) QCD corrections \cite{H2gQCD2} including finite top mass
effects \cite{H2gQCD3}; soft-gluon resummation effects \cite{softglu};
the first-order electroweak (EW) corrections \cite{DjG,ABDV0,APSU};
estimates of the next-to-next-to-next-to-leading order (NNNLO) QCD
corrections~\cite{Moch:2005ky} and of the mixed QCD-EW
corrections~\cite{qcdew}.

The Higgs sector of the Minimal Supersymmetric Standard Model (MSSM)
consists of two $SU(2)$ doublets, $H_1$ and $H_2$, whose relative
contribution to electroweak symmetry breaking is determined by the
ratio of vacuum expectation values of their neutral components,
$\tb\equiv v_2/v_1$. The spectrum of physical Higgs bosons is richer
than in the SM, consisting of two neutral CP-even bosons, $h$ and $H$,
one neutral CP-odd boson, $A$, and two charged bosons, $H^\pm$. The
couplings of the MSSM Higgs bosons to matter fermions differ from
those of the SM Higgs, and they can be considerably enhanced (or
suppressed) depending on $\tb$. As in the SM, gluon fusion is one of
the most important production mechanisms for the neutral Higgs bosons,
whose couplings to the gluons are mediated by top and bottom quarks
and their supersymmetric partners, the stop and sbottom squarks.

In the MSSM, the cross section for Higgs boson production in gluon
fusion is currently known at the NLO. The contributions arising from
diagrams with quarks and gluons, with full dependence on the Higgs and
quark masses, can be obtained from the corresponding SM results
\cite{SDGZ,HK,babis1,ABDV,BDV} with an appropriate rescaling of the
Higgs-quark couplings. The contributions arising from diagrams with
squarks and gluons were first computed under the approximation of
vanishing Higgs mass in ref.~\cite{Dawson:1996xz}, and the full
Higgs-mass dependence was included in later calculations
\cite{babis1,ABDV,BDV,MS}.  The contributions of two-loop diagrams
involving top, stop and gluino to both scalar and pseudoscalar Higgs
production were computed in the vanishing-Higgs-mass limit (VHML) in
refs.~\cite{HS,Franziska}, whose results were later confirmed and cast
in a compact analytic form in refs.~\cite{DS1,DDVS}. Finally, first
results for the NNLO contributions in the limit of vanishing Higgs
mass and degenerate stop and gluino masses were presented in
ref.~\cite{NNLOSUSY}.

The VHML can provide reasonably accurate results as long as the Higgs
mass is well below the threshold for creation of the massive particles
running in the loops. For the production of the lightest scalar Higgs,
this condition does apply to the two-loop diagrams involving top, stop
and gluino, but it obviously does not apply to the corresponding
diagrams involving the bottom quark, whose contribution can be
relevant for large values of $\tb$. In turn, the masses of the
heaviest scalar $H$ and of the pseudoscalar $A$ might very well
approach (or exceed) the threshold for creation of top quarks or even
of squarks. Unfortunately, retaining the full dependence on the Higgs
mass in the quark-squark-gluino contributions has proved a rather
daunting task. A calculation based on a combination of analytic and
numerical methods was presented in ref.~\cite{babis2} (see also
ref.~\cite{spiraDb}), but neither explicit analytic results nor a
public computer code have been made available so far.

However, results from the first year of supersymmetry (SUSY) searches
at the LHC (see, e.g., ref.~\cite{LHCSUSY}) set preliminary lower
bounds on the squark and gluino masses of the order of the TeV, albeit
for specific models of SUSY breaking. This suggests that -- if the
MSSM is actually realized in nature -- there might be wide regions of
its parameter space in which all three of the neutral Higgs bosons are
somewhat lighter than the squarks and the gluino. Approximate analytic
results for the quark-squark-gluino contributions can be derived in
this case. In particular, ref.~\cite{DS2} presented an approximate
evaluation of the bottom-sbottom-gluino contributions to scalar
production, based on an asymptotic expansion in the large
supersymmetric masses that is valid up to and including terms of
${\cal O}(\mb^2/m_\phi^2)$, ${\cal O}(\mb/\MS)$ and ${\cal
  O}(\mz^2/\MS^2)$, where $m_\phi$ denotes a Higgs boson mass and
$\MS$ denotes a generic superparticle mass. An independent calculation
of the quark-squark-gluino contributions to scalar production,
restricted to the limit of zero squark mixing and degenerate
superparticle masses, was also presented in ref.~\cite{hhm},
confirming the results of ref.~\cite{DS2} for the bottom
contributions. More recently, ref.~\cite{DDVS} presented an evaluation
of the quark-squark-gluino contributions to pseudoscalar production
that is also based on an asymptotic expansion in the large
supersymmetric masses, but does not assume any hierarchy between the
pseudoscalar mass and the quark mass, thus covering both the
top-stop-gluino and bottom-sbottom-gluino cases.

Exploiting the asymptotic-expansion techniques developed in
refs.~\cite{DS2} and \cite{DDVS}, we provide in this paper an
evaluation of the two-loop top-stop-gluino contributions to
Higgs-scalar production valid when the Higgs and the top quark are
lighter than stops and gluino, without assuming a specific hierarchy
between the Higgs mass and the top mass. In particular, we provide
explicit and compact analytic formulae which include terms up to
${\cal O}(m_\phi^2/\MS^2)$, ${\cal O}(\mt^2/\MS^2)$ and ${\cal
  O}(\mz^2/\MS^2)$. The results presented in this paper complement the
earlier ones of ref.~\cite{DS1}, which, being obtained via a Taylor
expansion in the Higgs mass, are not accurate for a Higgs mass
comparable to (or greater than) the top mass, as might well be the
case for the heaviest Higgs scalar of the MSSM. Our formulae can be
easily implemented in computer codes\footnote{An implementation of the
  MSSM gluon-fusion cross section in the {\tt POWHEG} framework was
  presented in ref.~\cite{BDSV}.}, allowing for an efficient and
accurate determination of the Higgs-boson production cross section in
the MSSM.

The paper is organized as follows: in section \ref{sec:general} we
summarize general results on the form factors for Higgs boson
production via gluon fusion in the MSSM. Section \ref{sec:2loopres}
contains our explicit results for the contributions arising from
two-loop top-stop-gluino diagrams, as well as a discussion of the
renormalization conditions for the parameters in the top/stop
sector. In section \ref{sec:num} we compare numerically the results of
our asymptotic expansion in the heavy masses with the results of a
Taylor expansion in the Higgs mass, up to and including terms of
${\cal O}(m_\phi^2/\mt^2)$ and ${\cal O}(m_\phi^2/\MS^2)$, discussing
the regions of applicability of the two different expansions and the
effect of different renormalization conditions. Finally, in the last
section we present our conclusions.

\vfill
\newpage

\section{Higgs boson production via gluon fusion in the MSSM}
\label{sec:general}

In this section we recall for completeness some general results on
Higgs boson production via gluon fusion in the MSSM. The leading-order
(LO) partonic cross section for the $gg\rightarrow \phi$ process (with
$\phi = h,H$) reads
\be
\sigma^{(0)}  =  
\frac{G_\mu \,\alpha_s^2 (\muR)  }{128\, \sqrt{2} \, \pi}\,
\left|{\cal H}^{1\ell}_\phi\right|^2~,
\label{ggh}
\ee
where $G_\mu$ is the muon decay constant, $\alpha_s(\muR)$ is the
strong gauge coupling expressed in the $\msbar$ renormalization scheme
at the scale $\muR$, and ${\cal H}_\phi$ is the form factor for the
coupling of the CP-even Higgs boson $\phi$ with two gluons, which we 
decompose in one- and two-loop parts as
\be
{\mathcal H}_\phi ~=~ {\mathcal H}_\phi^{1\ell}
~+~ \frac{\alpha_s}{\pi} \,  {\mathcal H}_\phi^{2\ell}
~+~{\cal O}(\alpha_s^2)~.
\label{Hdec}
\ee

The form factors for the lightest and heaviest Higgs mass eigenstates
can be decomposed as
\be
{\cal H}_{h} ~=~ T_F\,\left(-\sin\alpha \,{\mathcal H}_1 +
\cos\alpha \,{\mathcal H}_2 \right)~,~~~~~~
{\cal H}_{H} ~=~ T_F\,\left(\cos\alpha \,{\mathcal H}_1 +
\sin\alpha \,{\mathcal H}_2 \right)~,
\ee
where $T_F = 1/2$ is a color factor, $\alpha$ is the mixing angle in
the CP-even Higgs sector of the MSSM and ${\mathcal H}_i$ ($i = 1,2$)
are the form factors for the coupling of the neutral, CP-even
component of the Higgs doublet $H_i$ with two gluons. 
Focusing on the contributions involving the
third-generation quarks and squarks, and exploiting the structure of
the Higgs-quark-quark and Higgs-squark-squark couplings, we can write
to all orders in the strong interactions~\cite{DS1}
\bea
{\mathcal H}_1 & = & \lambda_t \,\left[
\mt \,\mu\,\sdt\,F_t
\,+ \mz^2 \,\sdbeta \,D_t  \right] \;+ 
\lambda_b \,\left[\mb\,A_b\,\sdb\,F_b \,+  2\,\mb^2\,G_b \,+ 
2\, \mz^2 \,c_\beta^2 \,D_b
\right]\,, \label{eq:H1} \\
{\mathcal H}_2 & = & \lambda_b\,\left[
\mb \,\mu\,\sdb\,F_b
\,-\mz^2 \,\sdbeta \,D_b  \right] + 
\lambda_t\, \left[
\mt\,A_t\,\sdt\,F_t \,+  2\,\mt^2\,G_t \,- 
2\, \mz^2 \,s_\beta^2 \,D_t
\right]\label{eq:H2}~.
\eea
In the equations above $\lambda_t = 1/\sin\beta$ and $\lambda_b =
1/\cos\beta$. Also, $\mu$ is the higgsino mass parameter in the MSSM
superpotential, $A_q$ (for $q=t,b$) are the soft SUSY-breaking
Higgs-squark-squark couplings and $\theta_q$ are the left-right squark
mixing angles (here and thereafter we use the notation $s_\varphi
\equiv \sin\varphi, \, c_\varphi \equiv \cos\varphi$ for a generic
angle $\varphi$). The functions $F_q$ and $G_q$ appearing in
eqs.~(\ref{eq:H1}) and (\ref{eq:H2}) denote the contributions
controlled by the third-generation Yukawa couplings, while $D_q$
denotes the contribution controlled by the electroweak, D-term-induced
Higgs-squark-squark couplings. The latter can be decomposed as
\be
D_q  = \frac{I_{3q}}2 \, \widetilde{G}_q
+ c_{2\theta_{\tilde q}} \, 
\left(\frac{I_{3q}}2 -  Q_q \,s^2_{\theta_\smallw} 
\right) \,\widetilde{F}_q \, , \label{eq:Dq}
\ee
where $I_{3q}$ denotes the third component of the electroweak isospin
of the quark $q$, $Q_q$ is the electric charge and $\theta_\smallw$ is
the Weinberg angle.

The form factors ${\mathcal H}_i$ can in turn be decomposed in one-
and two-loop parts as in eq.~(\ref{Hdec}).  The one-loop parts, ${\cal
  H}_i^{1\ell}$, contain contributions from diagrams involving quarks
($q$) or squarks ($\tilde q_i$). The functions entering ${\mathcal
  H}_i^{1\ell}$ are

\bea
F_q^{1\ell} ~=~ \widetilde F_q^{1\ell}
& =&  \frac{1}{2}\,\left[
\frac1{m^2_{\tilde{q}_{1}}} {\mathcal G}^{1\ell}_{0}(\tau_{\tilde{q}_{1}}) -
\frac1{m^2_{\tilde{q}_{2}}} {\mathcal G}^{1\ell}_{0}(\tau_{\tilde{q}_{2}}) 
\right]\, , \label{eq:F1l}\\
G_q^{1\ell} & =& \frac{1}{2}\,\left[
\frac1{m^2_{\tilde{q}_{1}}} {\mathcal G}^{1\ell}_{0}(\tau_{\tilde{q}_{1}}) +
\frac1{m^2_{\tilde{q}_{2}}} {\mathcal G}^{1\ell}_{0} (\tau_{\tilde{q}_{2}}) + 
\frac1{m_q^2} {\mathcal G}^{1\ell}_{1/2} (\tau_q)\right]~, \label{eq:G1l}\\
\widetilde G_q^{1\ell} & =& \frac{1}{2}\,\left[
\frac1{m^2_{\tilde{q}_{1}}} {\mathcal G}^{1\ell}_{0} (\tau_{\tilde{q}_{1}}) +
\frac1{m^2_{\tilde{q}_{2}}} {\mathcal G}^{1\ell}_{0} (\tau_{\tilde{q}_{2}}) 
\label{eq:Gtilde}\right]~,
\eea
where $\tau_k \equiv 4\,m_k^2/m_h^2$, and the functions ${\mathcal
G}^{1\ell}_{0}$ and ${\mathcal G}^{1\ell}_{1/2}$ read
\bea
{\mathcal G}^{1\ell}_{0} (\tau) & =& ~~~~\,\tau \!\left[ 1 + \frac{\tau}{4}\, 
 \ln^2 \left(\frac{\sqrt{1- \tau} - 1}{\sqrt{1- \tau} + 1}\right) \right]\,,
\label{eq:4} \\
{\mathcal G}^{1\ell}_{1/2} (\tau) & = & - 2\,\tau
 \left[ 1 - \frac{ 1 -\tau}4  \,   
 \ln^2 
  \left(\frac{\sqrt{1-\tau} - 1}{\sqrt{1-\tau} + 1} \right) \right] \,.
\label{eq:3}
\eea
The analytic continuations are obtained with the replacement $m_h^2
\rightarrow m_h^2 + i \epsilon$~. We remark that in the limit in which
the Higgs boson mass is much smaller than the mass of the particle
running in the loop, i.e.~$\tau\gg 1$, the functions ${\mathcal
  G}^{1\ell}_{0}$ and ${\mathcal G}^{1\ell}_{1/2}$ behave as
\be
{\mathcal G}^{1\ell}_{0} \rightarrow -\frac13 -\frac{8}{45\,\tau} 
~+~{\cal O}(\tau^{-2})~,~~~~~~~~~~
{\mathcal G}^{1\ell}_{1/2} \rightarrow -\frac43-\frac{14}{45\,\tau} 
~+~{\cal O}(\tau^{-2})~.
\label{Glimit}
\ee

The two-loop parts of the form factors, ${\cal H}_i^{2\ell}$, contain
contributions from diagrams involving quarks, squarks, gluons and
gluinos.
We point the reader to, e.g., section 2 of ref.~\cite{DS1} for
explicit formulae showing how ${\cal H}_i^{2\ell}$ (or, equivalently,
${\cal H}_\phi^{2\ell}$) enter the total NLO cross section for Higgs
boson production in hadronic collisions.
In the next section we present our new evaluation of the top/stop
contributions to ${\cal H}_i^{2\ell}$, based on an asymptotic
expansion in the stop and gluino masses.


\section{Two-loop contributions to the Higgs-production form factors}
\label{sec:2loopres}
In the case of the lightest Higgs boson $h$, the top/stop
contributions to the two-loop form factor $\Hhd$ are well under
control. Typically, the mass ratios between the lightest Higgs and the
particles running in the loops allow for the evaluation of the
relevant diagrams via a Taylor expansion in the Higgs mass, with the
zero-order term in the series -- for which ref.~\cite{DS1} provides
explicit analytic formulae -- already a good approximation to the full
result. In the case of the heaviest Higgs boson $H$, on the other
hand, the assumption that it is much lighter than the particles
running in the loops is valid only in a limited portion of the MSSM
parameter space. In particular, $\mH$ might very well sit around or
above the threshold for the creation of a real top-quark pair in the
loops, in which case -- as found in ref.~\cite{DDVS} for the
pseudoscalar -- a Taylor expansion in $\mH^2$ would certainly fail to
approximate the correct result for the Higgs-production form
factor. To address this possibility, we present in this section
explicit analytic results for the two-loop top/stop contributions to
the form factors ${\cal H}_i^{2\ell}$ that include terms up to ${\cal
  O}(m_\phi^2/\MS^2)$, ${\cal O}(\mz^2/\MS^2)$ and ${\cal
  O}(\mt^2/\MS^2)$, without assuming a specific hierarchy between the
Higgs mass and the top mass.

\begin{figure}[t]
\begin{center}
\mbox{
\epsfig{figure=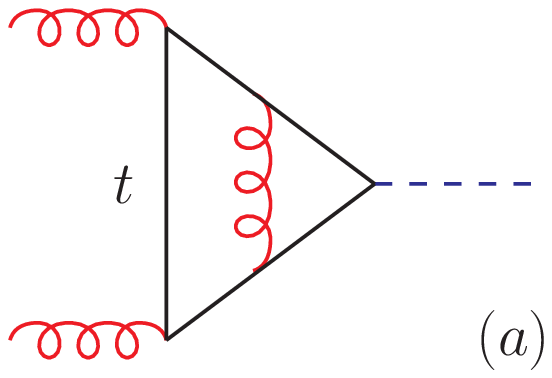,width=4.3cm}~~~~~~
\epsfig{figure=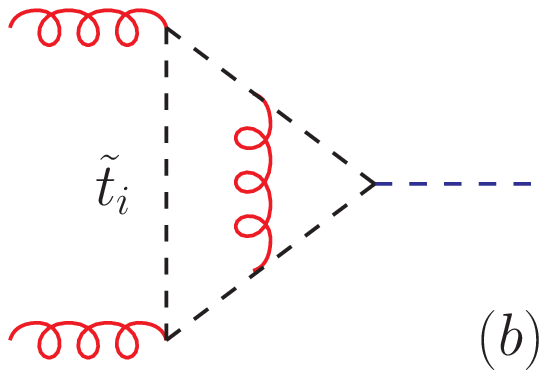,width=4.3cm}~~~~~~
\epsfig{figure=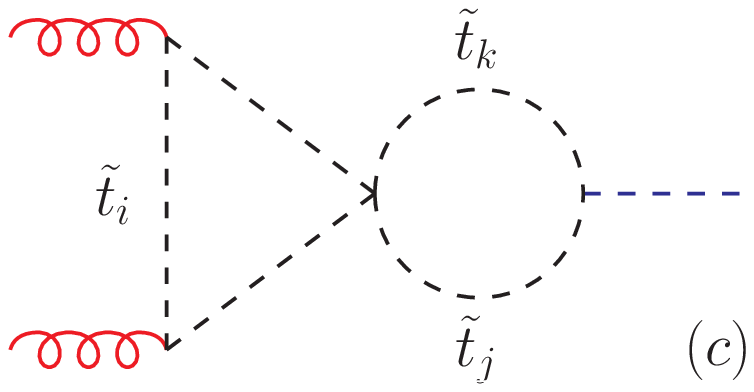,width=5.7cm}
}
\caption{Examples of two-loop diagrams for $gg\rightarrow\phi$ that do
  not involve gluinos.}
\label{fig:others}
\end{center}
\end{figure}

\begin{figure}[t]
\begin{center}
\mbox{
\epsfig{figure=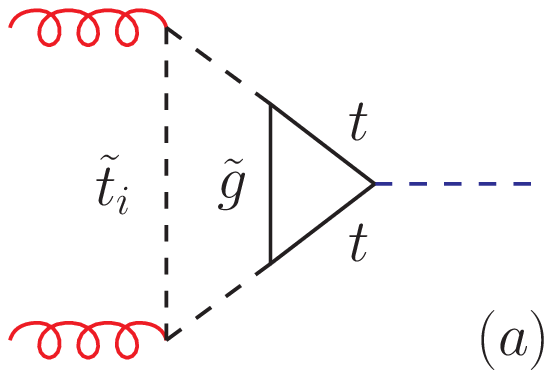,width=4.3cm}~~~~~~
\epsfig{figure=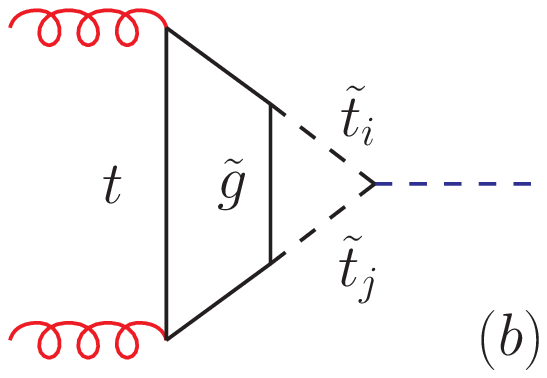,width=4.3cm}
}
\caption{Examples of two-loop diagrams for $gg\rightarrow\phi$
  involving gluinos.}
\label{fig:gluino}
\end{center}
\end{figure}

The top/stop contributions to ${\cal H}_i^{2\ell}$ come from two-loop
diagrams such as the ones depicted in figs.~\ref{fig:others} and
\ref{fig:gluino}. In analogy to what was done in refs.~\cite{DS1,DS2},
we can decompose the functions $F_t^{2\ell},\,G_t^{2\ell},\,
\widetilde F_t^{2\ell}$ and $\widetilde G_t^{2\ell}$ entering the
two-loop parts of eqs.~(\ref{eq:H1}) and (\ref{eq:H2}) as
\bea
F_t^{2\ell} &=& Y_{\stu} - Y_{\std} -\frac{4\,\cdt^2}{\tu-\td}\, Y_{\cdt^2}
\label{Fderiv}\,,\\
G_t^{2\ell} &=& Y_{\stu} + Y_{\std} + Y_t
\label{Gderiv}\,,\\
\widetilde F_t^{2\ell} &=& Y_{\stu} - Y_{\std}
+\frac{4\,\sdt^2}{\tu-\td}\,Y_{ \cdt^2}\,,
\label{Ftderiv} \\
\widetilde G_t^{2\ell} &=& Y_{\stu} + Y_{\std}\,.
\label{Gtderiv}
\eea

The various terms in eqs.~(\ref{Fderiv})--(\ref{Gtderiv}) can be split
in the contributions coming from diagrams with (s)top (s)quarks and
gluons ($g$, figs.~\ref{fig:others}a and \ref{fig:others}b); with a
quartic stop coupling ($4\tilde t$, fig.~\ref{fig:others}c); with top
quarks, stop squarks and gluinos ($\gl$, figs.~\ref{fig:gluino}a and
\ref{fig:gluino}b):
\be
Y_x = Y_x^{g}  + Y_x^{4\tilde t} + Y_x^{\gl}~~~~~~~(x=t,\stu,\std,\cdt^2)~.
\label{Ycon}
\ee
Furthermore, we remark that the term $Y_t$ entering eq.~(\ref{Gderiv})
contains only contributions from diagrams with a Higgs-top coupling,
figs.~\ref{fig:others}a and \ref{fig:gluino}a, therefore $Y_t^{4\tilde
  t} =0$. On the other hand, the terms $Y_{\stu},\, Y_{\std}$ and
$Y_{\cdt^2}$ in eqs.~(\ref{Fderiv})--(\ref{Gtderiv}) contain only
contributions from diagrams with a Higgs-stop coupling,
figs.~\ref{fig:others}b, \ref{fig:others}c, and \ref{fig:gluino}b.

\subsection{Top-gluon, stop-gluon and four-stop contributions}
\label{sec:others}
The top-gluon, stop-gluon and four-stop contributions to the terms
$Y_x$ in eq.~(\ref{Ycon}) can be extracted from the existing
literature, and we collect them in this section for completeness. We
assume that the parameters entering the one-loop parts of the form
factors ${\cal H}_i$ in eqs.~(\ref{eq:H1}) and (\ref{eq:H2}) are
expressed in the $\drbar$ renormalization scheme at the scale $Q$.

The contribution to the term $Y_t$ arising from two-loop diagrams with
top quarks and gluons (fig.~\ref{fig:others}a) must be computed for
arbitrary values of $\tau_t \equiv 4\mt^2/m_\phi^2\,$. It reads:
\be
2\,\mt^2\,Y^{g}_t ~=~ C_F\,\left[{\cal F}^{(2\ell,a)}_{1/2}(x_t) ~+~
{\cal F}^{(2\ell,b)}_{1/2}(x_t)\,\left(\ln\frac{\mt^2}{Q^2}-
\frac13\right)\right]~+~ C_A\; {\cal G}^{(2\ell,C_A)}_{1/2}(x_t)~,
\label{Ytglu}
\ee
where $C_F=4/3$ and $C_A=3$ are color factors, and exact expressions
for ${\cal F}^{(2\ell,a)}_{1/2}$, ${\cal F}^{(2\ell,b)}_{1/2}$ and
${\cal G}^{(2\ell,C_A)}_{1/2}\,$ as functions of $x_t \equiv
(\sqrt{1-\tau_t}-1)/(\sqrt{1-\tau_t}+1)$ are given in eqs.~(2.12),
(2.13) and (3.8) of ref.~\cite{ABDV}, respectively.

The contributions to the terms $Y_{\stu},\, Y_{\std}$ and $Y_{\cdt^2}$
arising from two-loop diagrams with stop squarks and gluons
(fig.~\ref{fig:others}b) and from diagrams with a quartic stop
coupling (fig.~\ref{fig:others}c) can, to the accuracy required by our
expansion, be computed in the limit of vanishing $m_\phi$. They
read~\cite{DS1}
\bea
\label{eq:glusqua}
Y_{\stu}^{g} &=& -\frac{1}{2\,\tu}
\,\left(\frac{3\,C_F}{4} + \frac{C_A}{6}\right)~,\\
\label{eq:quasqua}
Y_{\stu}^{4 \tilde t}&=&-\frac{C_F}{24}\,\left[
\frac{\cdt^2\,\tu+\sdt^2\,\td}{\tuq}
+\frac{\sdt^2}{\tuq\,\td}\,\left(
\tuq\,\ln\frac{\tu}{Q^2}-\tdq\,\ln\frac{\td}{Q^2}\right)\right]~,\\
\label{eq:quamix}
Y_{\cdt^2}^{4 \tilde t}&=&-\frac{C_F}{24}
\,\left[\frac{(\tu-\td)^2}{\tu\,\td}
-\frac{\tu-\td}{\td}\,\ln\frac{\tu}{Q^2}
-\frac{\td-\tu}{\tu}\,\ln\frac{\td}{Q^2}\right]~.
\eea
The term $Y_{\cdt^2}^{g}$ is zero, while the terms $Y_{\std}^{g}$ and
$Y_{\std}^{4 \tilde t}$ can be obtained by performing the
substitutions $\stu \leftrightarrow\std$ in eqs.~(\ref{eq:glusqua})
and (\ref{eq:quasqua}), respectively.

\subsection{Top-stop-gluino contributions}
\label{sec:gluino}

In this section we present our original results for the asymptotic
expansion of the top-stop-gluino contributions in the stop and gluino
masses. We retain in our formulae only terms that contribute to the
form factors ${\cal H}_i$ up to ${\cal O}(\mt^2/\MS^2)$, ${\cal
  O}(m_\phi^2/\MS^2)$ or ${\cal O}(\mz^2/\MS^2)$, where $\MS$ denotes
a generic superparticle mass.
Again, we assume that the one-loop parts of ${\cal H}_i$ in
eqs.~(\ref{eq:H1}) and (\ref{eq:H2}) are expressed in terms of
$\drbar$-renormalized parameters evaluated at the scale $Q$.
The top-stop-gluino contributions to the term $Y_t$, arising from
diagrams with a Higgs-top coupling (fig.~\ref{fig:gluino}a), read
\bea
\label{Ytgluino}
2\,\mt^2\,Y_t^{\gl} &=&
\frac{4}{3}\, {\cal F}^{(2\ell,b)}_{1/2}(\tau_t) \,
\frac{\delta m_t}{m_t}^{\!\scriptscriptstyle SUSY}
- ~\frac{C_F}4 \, {\mathcal G}^{1\ell}_{1/2} (\tau_t)\,\frac{\mg}{\mt}\,\sdt  \,
\left(\frac{x_1}{1-x_1} \ln x_1-\frac{x_2}{1-x_2}\ln x_2\right)\nn\\
&&+~ \sdt \frac{\mt}{\mg} \,{\cal R}_1 ~+~ \frac{\mt^2}{\mg^2}\,{\cal R}_2~,
\eea
where $x_i = \ti /\mg^2\,$ ($i=1,2$), and $\delta
m_t^{\scriptscriptstyle SUSY}$ denotes the SUSY contribution to the
top self-energy, in units of $\alpha_s/\pi$, expanded in powers of
$\mt$ up to terms of ${\cal O}(\mt^3)$
\bea
\label{dmtsusy}
\delta m_t^{\scriptscriptstyle SUSY} &=& 
-\frac{C_F}{4}\,\mt\,\left[
\sdt\,\frac{\mg}{\mt}\,\frac{\x1g}{1-\x1g} \ln \x1g
+ \frac12 \ln\frac{\mg^2}{Q^2}  +  
\frac{x_1 - 3}{4\, (1-x_1)} + \frac{x_1\,(x_1-2)}{2\,(1-x_1)^2} \,\ln x_1
\right.\nn\\
&+&\!\left.
\frac{\sdt\,\mt}{2\,\mg\,(1-x_1)^3} \left( 1-x_1^2 + 2\,x_1 \ln x_1 \right)
+\frac{\mt^2}{6\,\mg^2\,(1-x_1)^3} 
\left(x_1^2-5 x_1-2 -\frac{6\,x_1}{1-x_1} \ln x_1 \right)\right]\nn\\
&+&\!
 \biggr(x_1\longrightarrow x_2\,,~~\sdt \longrightarrow -\sdt\biggr)~.
\eea
The terms ${\cal R}_1$ and ${\cal R}_2$ in eq.~(\ref{Ytgluino})
collect contributions suppressed by $\mt/M$ and $\mt^2/M^2\,$,
respectively:
\bea
\label{eq:R1}
{\cal R}_1 &=&
\frac{C_A}{6 \left(1-x_1\right)^2} \,
\left[3\, \left(1-x_1+x_1 \ln x_1\right) \left(\ln \frac{\mt^2}{\mg^2}
-\mathcal{B}(\tau_t)
-\frac12\,{\mathcal{K}^{1\ell}(\tau_t)}+2\right) \right. \nn\\
& & ~~~~~~~~~~~~~~~~~+ \left. \vphantom{\frac{\mathcal{K}^{1\ell}}{2}} 
6\, x_1\, {\rm Li}_2\left(1-x_1\right) 
+2\, x_1+2\, x_1\left(1+x_1\right)\, \ln x_1-2\right] \nn\\
&-&\frac{C_F}{6\, x_1 \,\left(1-x_1\right)^3} 
\left[3\, \left(x_1-x_1^3+2\, x_1^2 \ln x_1\right) 
\left(\ln \frac{\mt^2}{\mg^2}-\mathcal{B}(\tau_t)
-\frac14\, {\mathcal{G}_{1/2}^{1\ell}(\tau_t)}
-\frac12\, {\mathcal{K}^{1\ell}(\tau_t)}+2\right) \right. \nn\\
& & \left. ~~~~~~~~~~~+\left(1-x_1\right)^3\, \ln \frac{\mg^2}{Q^2} 
+12\, x_1^2\, {\rm Li}_2\left(1-x_1\right)+5\, x_1^3-5\, x_1^2  
+x_1 -1 +2\, \left(x_1^3+2\, x_1^2\right) \ln x_1\right] \nn\\
&- &  \biggr(x_1 \longrightarrow x_2\biggr)~,\\
&&\nn\\
\label{eq:R2}
{\cal R}_2&=&
-\frac{C_A}{12 \left(1-x_1\right)^3} 
\left[3\, \left(1-x_1^2+2\, x_1 \ln x_1\right) 
\left(2\, \ln \frac{\mt^2}{\mg^2}-\mathcal{B}(\tau_t)
-\frac12\,{\mathcal{K}^{1\ell}(\tau_t)}+2\right) \right. \nn\\
&& \left. \vphantom{\frac{\mathcal{K}^{1\ell}}{2}} 
~~~~~~~~~~~~~~~~~~~~~
+24\, x_1 {\rm Li}_2\left(1-x_1\right) +1-x_1^2 
+2\, x_1\, \left(3\, x_1+10\right)\, \ln x_1\right] \nn\\
&+&  \frac{C_F}{18\, x_1\, \left(1-x_1\right)^4} 
\left\{ \vphantom{\frac{\mathcal{K}^{1\ell}}{2}} 
3\,x_1 \biggr[\left(1-x_1\right)\,(5\,x_1 - x_1^2+2)
+6\, x_1 \ln x_1\biggr] \times \right. \nn\\
&& ~~~~~~~~~~~~~~~~~~~~~\times \left(2\, \ln \frac{\mt^2}{\mg^2}
-\mathcal{B}(\tau_t) -\frac12\,{\mathcal{G}_{1/2}^{1\ell}(\tau_t)}
-\frac12\,{\mathcal{K}^{1\ell}(\tau_t)}+2\right) 
+6\, \left(1-x_1\right)^4 \ln \frac{m_{\tilde{g}}^2}{Q^2} \nn\\
&&  ~~~~~~~~~~~\left. \vphantom{\frac{\mathcal{K}^{1\ell}}{2}} 
+72\, x_1^2 \, {\rm Li}_2 (1-x_1) -x_1\, (1-x_1)^2\,(11\, x_1-26) 
-6\, \left(1-x_1\right)+6\, x_1^2 \left(2\, x_1+9\right)\, \ln x_1\right\} \nn\\
&+& \biggr(x_1 \longrightarrow x_2\biggr)~.
\eea
We recall that the function $\mathcal{G}^{1\ell}_{1/2}(\tau)$ is
defined in eq.~(\ref{eq:3}), while $\mathcal{B}(\tau)$ and
$\mathcal{K}^{1\ell}(\tau)$ are defined as
\be
\label{kappab}
{\cal B}(\tau) ~=~
2+\sqrt{1-\tau}\,\ln \left(\frac{\sqrt{1- \tau}-1}{\sqrt{1-\tau}+1}\right)~,
~~~~~~~~~~
{\cal K}^{1\ell}(\tau) ~=~ 
\frac{\tau}{2}\,\ln^2\left(\frac{\sqrt{1- \tau}-1}{\sqrt{1-\tau}+1}\right)~.
\ee
Finally, ${\cal F}^{(2\ell,b)}_{1/2}$ can be expressed directly as a
function of $\tau$ in terms of the other three functions:
\be
\label{F2lb} 
{\cal F}^{(2\ell,b)}_{1/2}(\tau) ~=~  -\frac32\,
\left[2 \,\mathcal{G}^{1\ell}_{1/2}(\tau) + \tau\,{\cal B}(\tau) 
-{\cal K}^{1\ell}(\tau)\right].
\ee

The top-stop-gluino contributions to the terms $Y_{\stu}$ and
$Y_{\cdt^2}$, arising from diagrams with a Higgs-stop coupling
(fig.~\ref{fig:gluino}b), read
\bea
\label{eq:Y1}
Y^{\gl}_{\stu} &=&
\left(\frac{C_F}{4}\,\frac{\sdt}{\mt\,\mg}\,\mathcal{G}^{1\ell}_{1/2}(\tau_t)
 - \frac{2\,C_F+C_A}{12\,\mg^2}\right)\,
\left(\frac{1}{1-x_1}+\frac{1}{\left(1-x_1\right)^2}\ln x_1\right)\nn\\
&+&\frac{C_F}{24\,\mg^2\,x_1^2\,(1-x_1)^3}\,
\left\{4\,(1-x_1)^3\,\left(1-\ln\frac{\mg^2}{Q^2}\right)
-3 \,x_1^2\,\mathcal{G}^{1\ell}_{1/2}(\tau_t)\,
\biggr[(1-x_1)(3-x_1) + 2\ln x_1\biggr]\right\}\nn\\
&+&\frac{C_F\,\sdt\,\mt}{6\,\mg^3\, x_1^2\, \left(1-x_1\right)^4} 
\left\{3 \,x_1^2\, \biggr[\left(1-x_1\right) 
\left(x_1+5\right)+2\,\left(2 \,x_1+1\right)\, \ln x_1\biggr] 
\left(\frac14\,{\mathcal{G}_{1/2}^{1\ell}(\tau_t)}
-\ln \frac{m_t^2}{\mg^2}\right) \right. \nn\\
&& ~~~~~~~~~~~~~~~~~~~~~~~~~
+\left(1-x_1\right)^4 \ln \frac{\mg^2}{Q^2} 
-12\, x_1^2\, \left(2 \,x_1+1\right) \,{\rm Li}_2\left(1-x_1\right) \nn\\ 
&& \left.\phantom{\frac14{\mathcal{G}_{1/2}^{1\ell}(\tau_t)}}
~~~~~~~~~~~~~~
-\left(1-x_1\right) \,\left(14\,x_1^2-3\,x_1+1\right) 
  -2\, x_1^2 \left(x_1^2+18\, x_1+5\right) \ln x_1\right\} \nn\\
&+&
 \frac{C_A\,\sdt\,\mt}{6\,\mg^3\,\left(1-x_1\right)^3}  
\left\{3\,\biggr[2-2\, x_1+\left(x_1+1\right)\, \ln x_1\biggr] 
\left(1+ \ln \frac{m_t^2}{\mg^2}\right)\right. \nn\\
&& ~~~~~~~~~~~~~~~~~~~~ \left. \vphantom{\frac{m_t^2}{\mg^2}} 
+6\, \left(1+x_1\right)\, {\rm Li}_2\left(1-x_1\right) 
+2\,x_1\,\left(1-x_1\right)+2\,\left(6\, x_1+1\right)\,\ln x_1\right\}\nn\\
&+&\frac{C_F\,\sdt\,m_\phi^2\, \mathcal{G}_{1/2}^{1\ell}(\tau_t)}
{48\,\mt\,\mg^3\, x_1\, \left(1-x_1\right)^4}
\, \biggr[\left(1-x_1\right) \left(x_1^2-5\, x_1-2\right)-6\, x_1 \ln x_1\biggr]~,\\
&&\nn\\
&&\nn\\
Y^{\gl}_{\cdt^2} &=& 
\label{eq:Yc}
-\frac{C_F\,\mg}{8\,\sdt\,\mt}\,\mathcal{G}_{1/2}^{1\ell}(\tau_t)\,
\left(\frac{x_1}{1-x_1}\,\ln x_1-\frac{x_2}{1-x_2}\,\ln x_2\right)\nn\\
&+&\!\!\left\{
\frac{C_F\,\mt}{12\,\sdt\,\mg\,x_1\left(1-x_1\right)^3}
\left[\,3\,x_1 \left(x_1^2-2\, x_1 \,\ln x_1-1\right) \!
\left(\frac14\,{\mathcal{G}_{1/2}^{1\ell}(\tau_t)}
-\ln \frac{m_t^2}{\mg^2}\right)
+\left(1-x_1\right)^3 \ln \frac{\mg^2}{Q^2} \right. \right.\nn\\
&& ~~~~~~~~~~~~~~~~~~~~~~~~~~~~~~~ 
\left. \vphantom{\frac{m_t^2}{\mg^2}} 
+12 \,x_1^2\, {\rm Li}_2\left(1-x_1\right)
-\left(1-2 x_1\right) \left(1-x_1\right)^2
+2 \,x_1^2\, \left(x_1+5\right) \ln x_1\right]\nn\\
&&+~  \frac{C_A\,\mt\,x_1}{12\,\sdt\,\mg\, \left(1-x_1\right)^2} 
\left[\left(x_1 - 1 - \ln x_1\right)
\left(3\, \ln \frac{\mt^2}{\mg^2}+1\right)
-6\, {\rm Li}_2\left(1-x_1\right)-2\,  \left(x_1+2\right) \ln x_1\right]\nn\\
&&+~\frac{C_F\,m_\phi^2\,\mathcal{G}_{1/2}^{1\ell}(\tau_t)}
     {32\,\sdt\,\mg\,\mt\left(1-x_1\right)^2}\,
\left[\frac{\left(1-x_1\right) \left(x_1+x_2-2 x_1 x_2\right)}
{\left(1-x_2\right) \left(x_1-x_2\right)} 
+\frac{2 x_1 \left(x_1^2+x_1 x_2-2 x_2\right) \ln x_1}
{\left(x_1-x_2\right)^2}\right] \nn\\
&&\nn\\
&&-~\biggr(x_1\longleftrightarrow x_2\biggr)~\biggr\}~,
\eea
while $Y^{\gl}_{\std}$ can be obtained by performing the substitutions
$x_1\rightarrow x_2$ and $\sdt \rightarrow -\sdt$ in
eq.~(\ref{eq:Y1}).

We remark that some care is required in order to properly include in
the form factors ${\cal H}_i$ only terms up to ${\cal
  O}(m_\phi^2/\MS^2)$, ${\cal O}(\mz^2/\MS^2)$ and ${\cal
  O}(\mt^2/\MS^2)$. In particular, in the calculation of the function
$F_t^{2\ell}$, eq.~(\ref{Fderiv}), we must use the full formulae for
$Y_{\stu}^{\gl}$ and $Y_{\cdt^2}^{\gl}$ in eqs.~(\ref{eq:Y1}) and
(\ref{eq:Yc}), respectively.  On the other hand, in the calculation of
the functions $G_t^{2\ell},\,\widetilde F_t^{2\ell}$ and $\widetilde
G_t^{2\ell}$, eqs.~(\ref{Gderiv})--(\ref{Gtderiv}), we must retain
only the terms in the first two lines of eq.~(\ref{eq:Y1}) for
$Y_{\stu}^{\gl}$, and only the term in the first line of
eq.~(\ref{eq:Yc}) for $Y_{\cdt^2}^{\gl}$.

As a first check of the correctness of our calculation, we verified
that by taking the VHML (i.e., taking $m_\phi\rightarrow 0$) in the
formulae presented in this section, which implies
$\mathcal{G}_{1/2}^{1\ell}(\tau_t) \rightarrow -4/3,~
\mathcal{K}^{1\ell}(\tau_t) \rightarrow -2,~
\mathcal{B}(\tau_t)\rightarrow 0$ and ${\cal F}^{(2\ell,b)}_{1/2}
\rightarrow 0$, we obtain for the top-stop-gluino contributions the
same result that we would obtain by expanding the VHML results of
ref.~\cite{DS1} in powers of $\mt$ up to and including ${\cal
  O}(\mt^2/M^2)$.
It is also straightforward to check that, by performing the trivial
replacement $t\rightarrow b$ in the formulae presented in this section
and then dropping all terms that contribute to the form factors beyond
${\cal O}(\mb^2/m_\phi^2)$, ${\cal O}(\mb/M)$ and ${\cal
  O}(\mz^2/M^2)$, we recover the results of ref.~\cite{DS2} for the
bottom-sbottom-gluino contributions. To this effect it must be kept in
mind that $\mathcal{G}_{1/2}^{1\ell}(\tau_b)$ and
$\mathcal{K}^{1\ell}(\tau_b)$ are of ${\cal O}(\mb^2/m_\phi^2)$, while
$\mathcal{B}(\tau_b) = 2 - \ln(-m_\phi^2/\mb^2) + {\cal
  O}(\mb^2/m_\phi^2)$.

\subsection{On-shell renormalization scheme for the stop parameters}
\label{sec:schemes}

If the parameters entering the one-loop part of the form factors are
expressed in a renormalization scheme different from $\drbar$, the
two-loop results presented in the previous section must be shifted in
a way analogous to that described in section 3.2 of ref.~\cite{DS1},
to which we point the reader for details. Indeed, in our ``on-shell''
(OS) scheme we adopt the same prescriptions as in ref.~\cite{DS1} for
the input parameters that are subject to ${\cal O}(\alpha_s)$
corrections: the top and stop masses are defined as the poles of the
corresponding propagators; the counterterm of the stop mixing angle
$\theta_t$ is chosen as to cancel the anti-hermitian part of the stop
wave-function renormalization matrix; the trilinear coupling $A_t$ is
treated as a derived quantity, related to the other parameters in the
top/stop sector by
\be
\label{mixing}
\sdt = \frac{2\,\mt\,(A_t + \mu\cot\beta)}{\tu-\td}~.
\ee

Some differences with respect to the treatment in ref.~\cite{DS1}
arise, however, due to the fact that the results presented in that
paper were obtained in the VHML for arbitrary values of the top mass,
while the results presented here are valid up to and including terms
of ${\cal O}(m_\phi^2/M^2)$ and ${\cal O}(\mt^2/M^2)$, without
assuming a hierarchy between $m_\phi$ and $\mt$. Defining the
$\drbar$--OS shift for a generic parameter $x$ according to
$x^{\drbar} = x^{\rm OS}+ (\alpha_s/\pi)\,\delta x$,
we need here to expand the various shifts in powers of $\mt$. Up to
the order relevant to our calculation, the explicit expressions for
the shifts in the stop masses and mixing angle read
\bea 
\label{dT1}
\delta \tu &=& 
\frac{C_F}{4}\,\tu\,\left\{ 3\, \ln \frac{\tu}{Q^2} - 3 -
 \cdt^2 \left( \ln \frac{\tu}{Q^2} -1 \right) 
-\sdt^2 \frac{\td}{\tu} \left( \ln \frac{\td}{Q^2} -1 \right)
 \right. \nn \\
& & ~~~~~~~~~~~~~ -6 \frac{\mg^2}{\tu} 
- 2 \left( 1 - 2  \frac{\mg^2}{\tu} \right) \ln \frac{\mg^2}{Q^2}
- 2 \left( 1-\frac{\mg^2}{\tu} \right)^2
\ln \left| 1-\frac{\tu}{\mg^2} \right|\nn\\
& & ~~~~~~~~~~~~~\left. -\frac{4\,\sdt\,\mt\,\mg}{\tu}
\left[ \ln \frac{\mg^2}{Q^2} + \left( 1-\frac{\mg^2}{\tu} \right)
\ln \left| 1-\frac{\tu}{\mg^2} \right| - 2 \right]\,\right\}~,\\
&&\nn\\
\label{dst}
\delta \theta_t &=& 
\frac{C_F}{4}\frac{\cdt\,\sdt}{(\tu-\td)}\,\left\{ 
\tu\,\left(\ln \frac{\tu}{Q^2}-1\right)
-\frac{2\,\mt\,\mg}{\sdt}\,\left[
\ln \frac{\mg^2}{Q^2} + \left( 1-\frac{\mg^2}{\tu} \right)
\ln \left| 1-\frac{\tu}{\mg^2} \right| - 2 \right]\,\right\}\nn\\
&&\nn\\
&+&\biggr(\stu\longleftrightarrow \std\,,
~~\sdt\longrightarrow-\sdt\,,~~\cdt\longrightarrow-\cdt\biggr)~, 
\eea
where the analogous expression for $\delta\td$ can be obtained by
performing the substitutions $\stu\leftrightarrow \std$ and $\sdt
\rightarrow -\sdt$ in eq.~(\ref{dT1}). We also define $\delta \sdt =
2\, \cdt\, \delta \theta_t$ and $\delta \cdt = - 2\, \sdt\, \delta
\theta_t$.  The shift for the top mass reads
\be
\label{dmt}
\delta \mt = 
\frac{C_F}{4}\,m_t\,\left(3 \ln\frac{\mt^2}{Q^2} - 5\right)
+ \delta m_t^{\scriptscriptstyle SUSY},
\ee
where the SUSY contribution $\delta m_t^{\scriptscriptstyle SUSY}$
was given in eq.~(\ref{dmtsusy}). Finally, the shift for the trilinear
coupling $A_t$ can be expressed in terms of the other shifts according
to
\be
\label{dAt}
\delta A_t ~=~ \left(\frac{\delta \tu-\delta\td}{\tu-\td}
+\frac{\delta\sdt}{\sdt}-\frac{\delta\mt}{\mt}\right)(A_t+\mu\,\cot\beta)~.
\ee

If the one-loop form factors are evaluated in terms of OS parameters,
the two-loop functions in eqs.~(\ref{Fderiv})--(\ref{Gtderiv}) must be
replaced by
\bea 
F_t^{2\ell} & \longrightarrow & F_t^{2\ell} ~+~
\frac16\,\left[
\frac{\delta\tu}{\tuq}-\frac{\delta\td}{\tdq}-\left(
\frac{\delta\mt}{\mt}+\frac{\delta\sdt}{\sdt}\right)
\,\left(\frac{1}{\tu}-\frac{1}{\td}\right)
- \frac{2 \,m_\phi^2}{15}\,\frac{\delta \mt}{\mt}
\left(\frac{1}{\tuq}-\frac{1}{\tdq}\right)
\right]\,,\nn\\
&&\label{shiftF}\\
\label{shiftG}
G_t^{2\ell} & \longrightarrow & G_t^{2\ell} ~+~
\frac16\,\left[
\frac{\delta\tu}{\tuq}+\frac{\delta\td}{\tdq}-
2\,\frac{\delta\mt}{\mt}\,\left(\frac{1}{\tu}+\frac{1}{\td}\right)\right]
~-~\frac23\, {\cal F}^{(2\ell,b)}_{1/2}(\tau_t) \, 
\frac{\delta m_t}{m_t^3}~,\\
&&\nn\\
\label{shiftFt}
\widetilde F_t^{2\ell} & \longrightarrow & \widetilde F_t^{2\ell} ~+~
\frac16\,\left[
\frac{\delta\tu}{\tuq}-\frac{\delta\td}{\tdq}
- \frac{\delta\cdt}{\cdt}\,\left(\frac{1}{\tu}-\frac{1}{\td}\right)\right]~,\\
&&\nn\\
\label{shiftGt}
\widetilde G_t^{2\ell} & \longrightarrow & \widetilde G_t^{2\ell} ~+~
\frac16\,\left[
\frac{\delta\tu}{\tuq}+\frac{\delta\td}{\tdq}\right]~.
\eea
In addition, the two-loop form factor ${\cal H}_2^{2\ell}$ gets
contributions originating from the shift in $A_t$:
\be
{\cal H}_2^{2\ell} \longrightarrow {\cal H}_2^{2\ell} ~-~
\frac{\mt\,\sdt}{6\,s_\beta}\,\left[
\delta A_t \,\left(\frac{1}{\tu}-\frac{1}{\td}\right)
+ \frac{2 \,m_\phi^2}{15}\,\delta A_t \,
\left(\frac{1}{\tuq}-\frac{1}{\tdq}\right)\right]
~.\label{shiftA}
\ee

Once again, care is required in order to properly include in the form
factors ${\cal H}_i$ only terms up to ${\cal O}(m_\phi^2/\MS^2)$,
${\cal O}(\mz^2/\MS^2)$ and ${\cal O}(\mt^2/\MS^2)$. In particular:
\begin{enumerate}\itemsep0pt
\item the shifts $\delta \tu$, $\delta \td$ and $\delta
  \theta_t$ must be computed up to ${\cal O}(\mt)$ in $F_t^{2\ell}$,
  eq.~(\ref{shiftF}), while they must be truncated at order zero in
  $\mt$ in $G_t^{2\ell}$, $\widetilde F_t^{2\ell}$ and $\widetilde
  G_t^{2\ell}$, eqs.~(\ref{shiftG})--(\ref{shiftGt});
\item in $F_t^{2\ell}$, eq.~(\ref{shiftF}), the first
  occurrence of $\delta \mt$ must be computed up to ${\cal O}(\mt^2)$,
  while the second occurrence, in the term proportional to $m_\phi^2$,
  must be truncated at order zero in $\mt$;
\item in $G_t^{2\ell}$, eq.~(\ref{shiftG}), the first
  occurrence of $\delta \mt$ must be truncated at ${\cal O}(\mt)$,
  while the second occurrence, in the term proportional to ${\cal
    F}^{(2\ell,b)}_{1/2}(\tau_t)$, must be computed up to ${\cal
    O}(\mt^3)$;
\item finally, in eq.~(\ref{shiftA}) the first occurrence of $\delta
  A_t$ must be computed up to ${\cal O}(\mt)$ by means of
  eq.~(\ref{dAt}), while the second occurrence, in the term
  proportional to $m_\phi^2$, must be truncated at ${\cal
    O}(\mt^{-1})$.
\end{enumerate}

Ref.~\cite{hhm} provides formulae for the two-loop SUSY contributions
to the form factors for scalar production in gluon fusion, in the OS
renormalization scheme, also based on an asymptotic expansion in the
superparticle masses but restricted to the limit of zero squark mixing
and degenerate superparticle masses. We checked that our OS results
agree with those of ref.~\cite{hhm} in the simplified limit considered
in that paper, after taking into account a difference in the overall
normalization factor and the fact that ref.~\cite{hhm} employs the
opposite convention for the sign of $\mu$ with respect to our
eq.~(\ref{mixing}).


\section{A numerical example}
\label{sec:num}

We will now discuss a numerical evaluation of the two-loop SUSY
contributions to the form factors for scalar Higgs production in a
representative region of the MSSM parameter space.

The SM parameters entering our calculation include the $Z$ boson mass $\mz 
= 91.1876$ GeV, the $W$ boson mass $\mw = 80.399$ GeV and the strong 
coupling constant $\alpha_s(\mz) = 0.118$ \cite{PDG}. For the pole mass of 
the top quark we take $M_t = 173.2$ GeV \cite{topmass}. For the relevant 
SUSY parameters we choose
\be 
\label{susypar} 
m_Q=m_U=\mu~=~ 1~{\rm TeV}\,,~~~A_t ~=~ 2~{\rm TeV}\,,~~~ 
\mg ~=~ 800~{\rm GeV}\,,~~~\tan\beta ~=~ 5~, 
\ee 
where $m_Q$ and $m_U$ are the soft SUSY-breaking masses for the left
and right stops, respectively. For a given value of the pseudoscalar
mass $\ma$, the scalar masses $m_h$ and $\mH$ and the mixing angle
$\alpha$ are computed including the leading one-loop corrections of
${\cal O}(\alpha_t)$ and the leading two-loop corrections of ${\cal
  O}(\alpha_s \alpha_t)$~\cite{DSZ}.

\begin{figure}[tp]
\begin{center}
\mbox{
\epsfig{figure=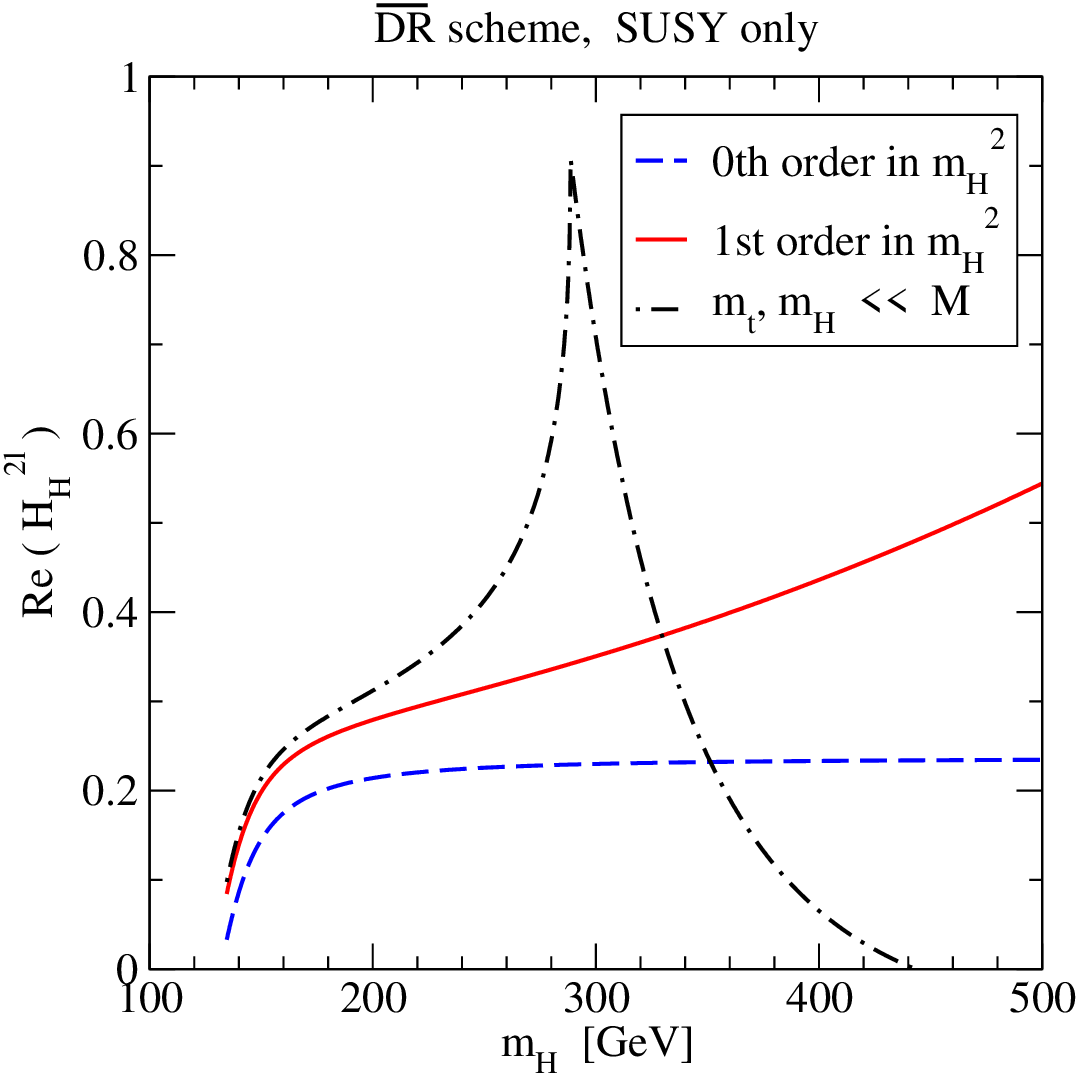,width=7.8cm}~~~~~
\epsfig{figure=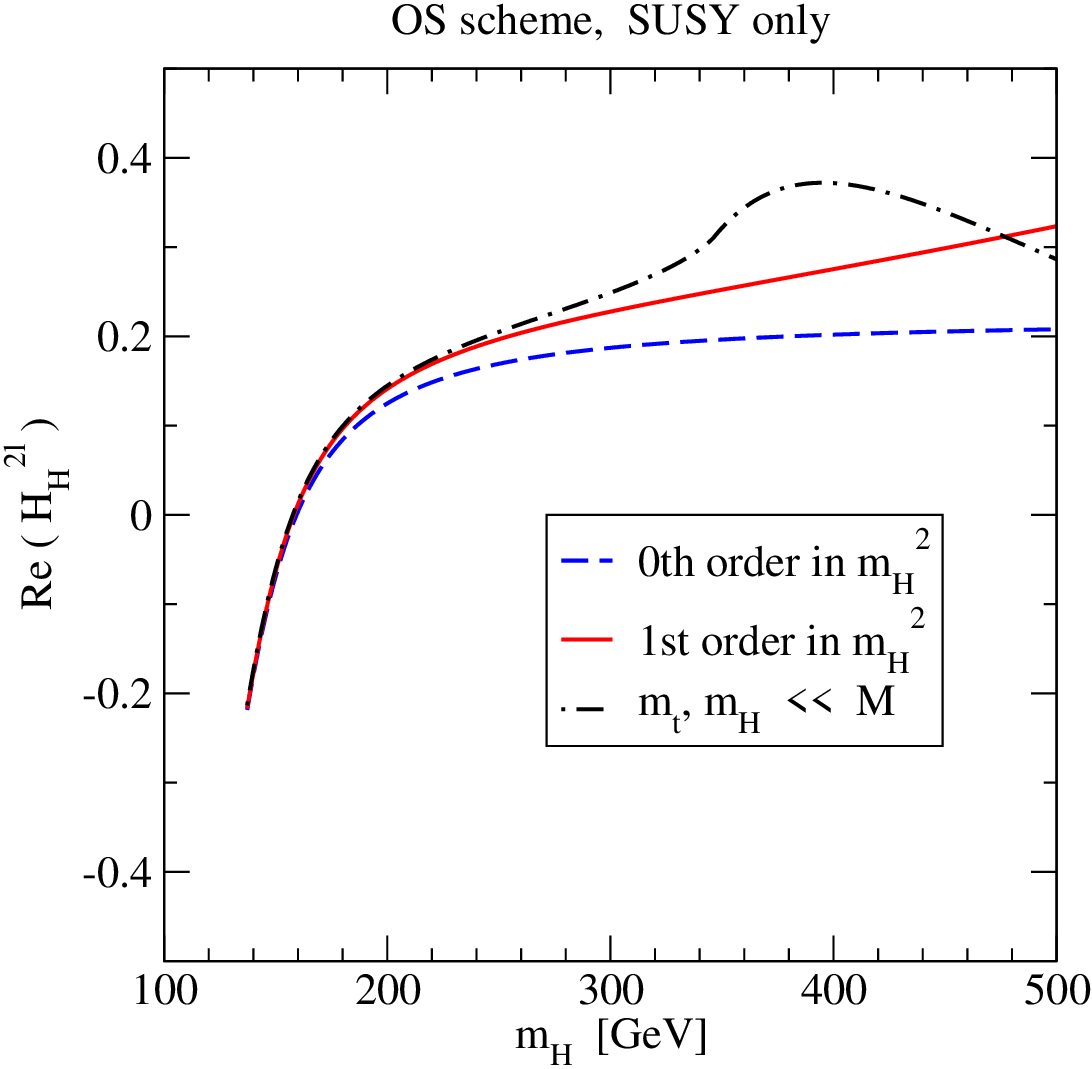,width=7.8cm}
}
\caption{Real part of the SUSY contributions to $\HHd$, plotted as a
  function of $\mH$. The choice of SUSY parameters and the meaning of
  the different curves are explained in the text. The plot on the left
  refers to the $\drbar$ scheme, while the plot on the right refers to
  the OS scheme.}
\label{fig:prodH}
\end{center}
\end{figure}

In fig.~\ref{fig:prodH} we show the real part of the SUSY (i.e., all
except top-gluon) contributions to the two-loop form factor for
heaviest-Higgs production, $\HHd$, as a function of $\mH$. Since, as
mentioned above, $\mH$ is not a free parameter in our calculation, its
variation is obtained by varying $\ma$ between 100 GeV and 500
GeV. For simplicity, in the computation of the form factor we
neglected the small D-term-induced electroweak contributions. The left
plot in figure \ref{fig:prodH} is obtained assuming that the
parameters $\mt,\,m_{\stu},\,m_{\std}$ and $\theta_t$ entering the
one-loop part of the form factor, $\HHu$, are expressed in the
$\drbar$ renormalization scheme at the scale $Q=1$ TeV. In this case
we extract the $\drbar$ top mass $\mt(Q)$ from the input value for the
pole mass $M_t$ by means of eq.~(B2) of ref.~\cite{DSZ}, and we
interpret the input parameters $m_Q,\,m_U$ and $A_t$ in
eq.~(\ref{susypar}) directly as running parameters evaluated at the
scale $Q$. The right plot, on the other hand, is obtained assuming
that the parameters $\mt,\,m_{\stu},\,m_{\std}$ and $\theta_t$
entering $\HHu$ are expressed in the OS scheme described in section
3.2 of ref.~\cite{DS1}. In this case we identify $\mt$ directly with
the pole mass $M_t$, and we interpret the input parameters $m_Q,\,m_U$
and $A_t$ in eq.~(\ref{susypar}) as the parameters that can be
obtained by rotating the diagonal matrix of the physical stop masses
by the ``physical'' angle $\theta_t$, defined through eq.~(37) of
ref.~\cite{DS1}.

In each plot, the dashed (blue) line represents the result obtained in
the VHML, as given in ref.~\cite{DS1}, while the solid (red) line
represents the result computed at the first order of a Taylor
expansion in $\mH^2$, i.e.~it includes the effect of terms of ${\cal
  O}(\mH^2/\mt^2)$ and ${\cal O}(\mH^2/M^2)$ which were also computed
in ref.~\cite{DS1} but proved too lengthy to be presented in analytic
form. The dot-dashed (black) line represents instead the result of the
asymptotic expansion in the superparticle masses derived in this
paper. The latter is applicable when both $\mt$ and $\mH$ are smaller
than the generic superparticle mass $M$, as is indeed the case here
since $M\approx1$ TeV, but it does not require any specific hierarchy
between $\mH$ and $\mt$.

The comparison between the dashed and solid lines shows that, as $\mH$
increases, the effect of the terms of ${\cal O}(\mH^2/\mt^2)$ and
${\cal O}(\mH^2/M^2)$ becomes more and more relevant, and the VHML
does not provide an accurate approximation to $\HHd$. Furthermore, the
comparison between the dot-dashed and solid lines shows that, even if
the inclusion of the first-order terms pushes the validity of the
Taylor expansion up to larger values of $\mH$, the Taylor expansion
fails anyway when $\mH$ gets close to the threshold for the production
of a real top-quark pair in the loops. In that case one can use the
result of our asymptotic expansion in $M$, provided that the latter is
still considerably larger than $\mH$.

A few additional comments are in order concerning the comparison
between the left ($\drbar$) and right (OS) plots in
fig.~\ref{fig:prodH}. There is no reason to expect the plots to look
similar to each other, first of all because the difference between the
values of $\HHd$ in the two schemes is compensated for, up to
higher-order terms, by a shift in the value of the one-loop form
factor, $\HHu$, and also because the different interpretation of the
input parameters in the two schemes means that, by using the numerical
inputs in eq.~(\ref{susypar}) for both schemes, we are in fact
considering two different points of the MSSM parameter space. This
said, a striking difference between the two schemes is visible in the
behavior of the asymptotic expansion (i.e., the dot-dashed line)
around the threshold for the production of a real top-quark pair in
the loops. The fact that in the $\drbar$ plot the threshold is located
at a lower value of $\mH$ than in the OS plot is an artifact, due to
lower value of the MSSM running top mass with respect to the pole top
mass (indeed, for our choice of parameters $\mt(Q) = 144.3$ GeV). The
much sharper behavior around the threshold of the dot-dashed line in
the $\drbar$ plot, on the other hand, can be traced back to the
contribution of the first term in the right-hand side of
eq.~(\ref{Ytgluino}) for $Y_t^{\gl}$. That term reflects the fact that
the running top mass of the MSSM (i.e., including the stop-gluino
contribution) is used in the top-quark contribution to $\HHu$, and it
is canceled out by the last term of eq.~(\ref{shiftG}) if the pole top
mass (or, for that matter, the running top mass of the SM) is used
instead. Indeed, we checked that, in a ``mixed'' renormalization
scheme in which the stop contributions to $\HHu$ are expressed in term
of running parameters (including the MSSM running top mass) but the
top-quark contribution is expressed in terms of the pole top mass, the
qualitative behavior of the dot-dashed line around the threshold would
be similar to the one in the OS plot.

\begin{figure}[tp]
\begin{center}
\mbox{
\epsfig{figure=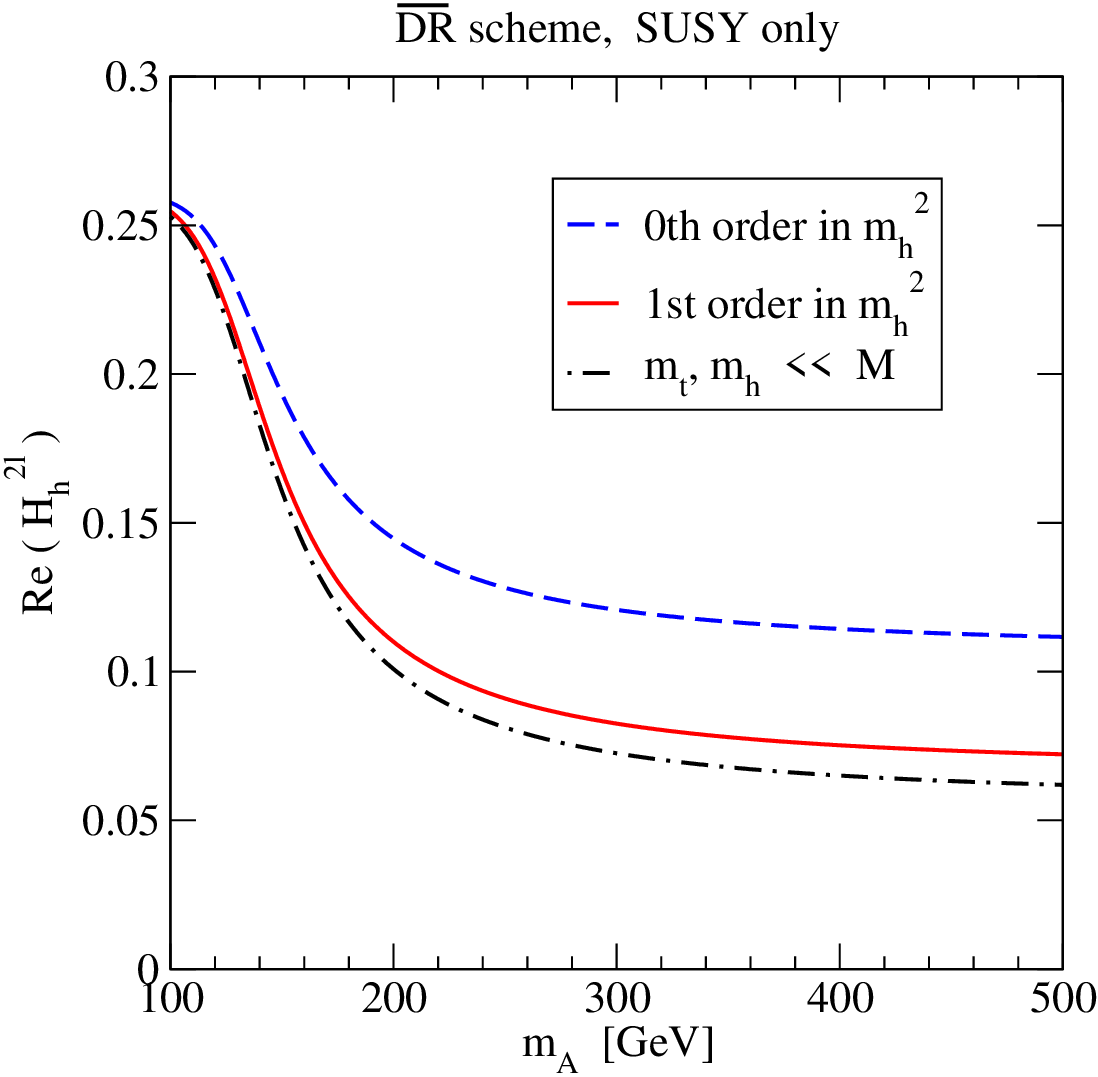,width=7.8cm}~~~~~~
\epsfig{figure=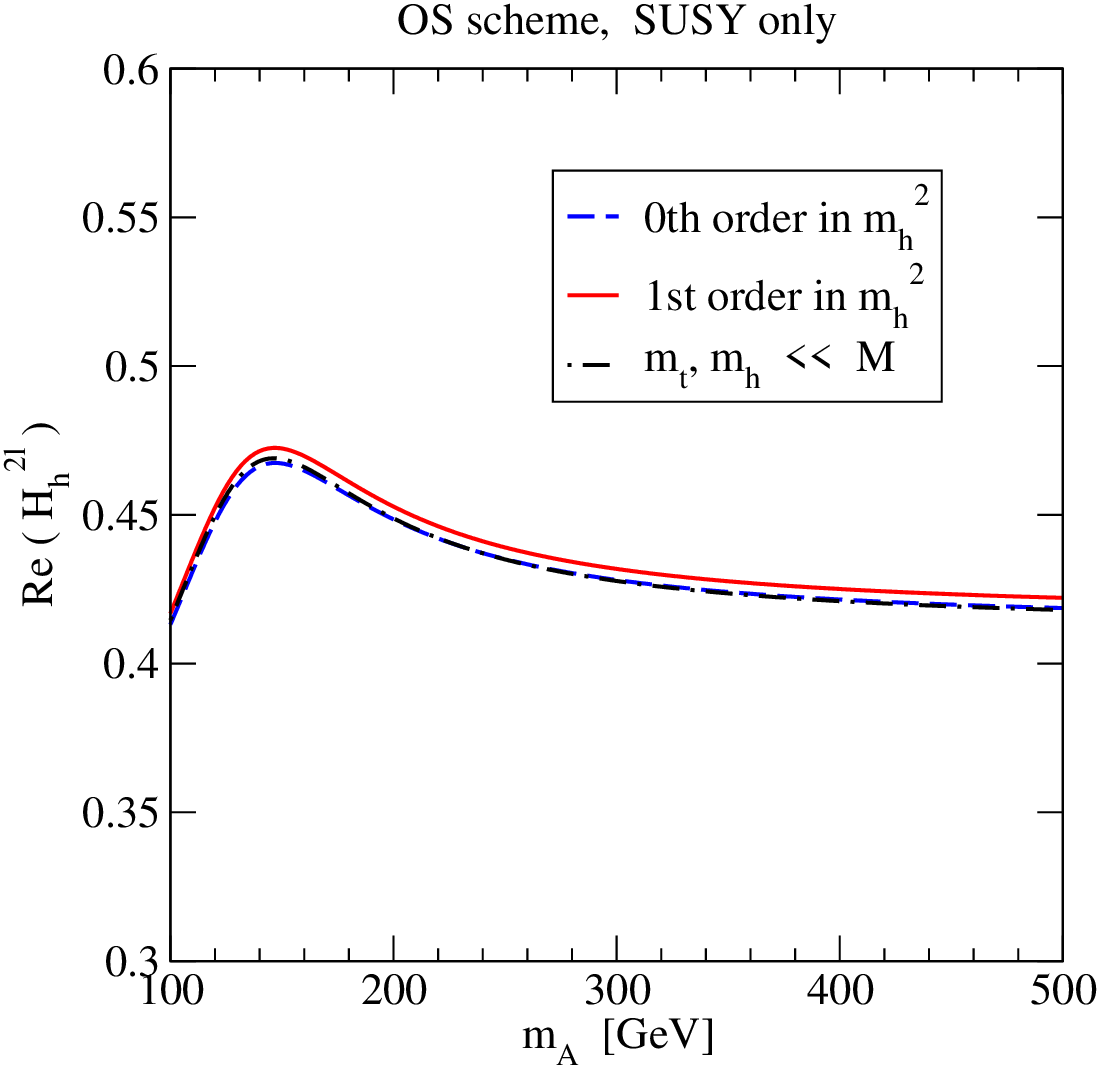,width=7.8cm}
}
\caption{Real part of the SUSY contributions to $\Hhd$, plotted as a
  function of $\ma$. The choice of SUSY parameters and the meaning of
  the different curves are explained in the text. The plot on the left
  refers to the $\drbar$ scheme, while the plot on the right refers to
  the OS scheme.}
\label{fig:prodh}
\end{center}
\end{figure}

To conclude this discussion, we show in fig.~\ref{fig:prodh} the real
part of the SUSY contributions to the two-loop form factor for
lightest-Higgs production, $\Hhd$, as a function of the pseudoscalar
mass $\ma$, which is varied in the same range used to produce
fig.~\ref{fig:prodH}. The meaning of the different curves is the same
as in fig.~\ref{fig:prodH}, and again the left plot is obtained
assuming that the parameters entering the one-loop form factor $\Hhu$
are expressed in the $\drbar$ scheme, while the right plot is obtained
assuming that they are expressed in the OS scheme.

In the MSSM the mass of the lightest Higgs scalar $h$ is bounded from
above, and for large enough values of the pseudoscalar mass it becomes
independent of $\ma$, as do the couplings of $h$ to the top quark and
to the stops. Indeed, for the choice of SUSY parameters in
eq.~(\ref{susypar}) our crude ${\cal O}(\alpha_t+\alpha_t\alpha_s)$
calculation of the Higgs mass yields $m_h<123.8$ GeV in the $\drbar$
plot and $m_h<122.5$ GeV in the OS plot, and all the curves in
fig.~\ref{fig:prodh} become essentially flat for $\ma> 250$ GeV. Due
to the relative smallness of $m_h$ no real-particle threshold is
crossed, thus the result of the asymptotic expansion (dot-dashed line)
is rather close to the result of the Taylor expansion at the first
order in $m_h^2$ (solid line).

However, a comparison between the left and right plots of
fig.~\ref{fig:prodh} shows that in the $\drbar$ calculation the VHML
result (dashed line) provides a less-than-perfect approximation to
$\Hhd$, while in the OS calculation the effect of the terms
proportional to $m_h^2$ is small, and the VHML result essentially
overlaps with the other two results. This difference between the two
schemes can again be traced to the contribution of the first term in
the right-hand side of eq.~(\ref{Ytgluino}), i.e.~to the choice of
renormalization scheme for the top mass entering the top-quark
contribution to $\Hhu$. Even in this case we checked that, in a
``mixed'' scheme in which the top-quark contribution to $\Hhu$ is
expressed in terms of the pole top mass while the stop contributions
are expressed in terms of running parameters, the VHML would provide
as good an approximation to $\Hhd$ as it does in the full OS scheme.

\section{Conclusions}
\label{sec:concl}

The calculation of the production cross section for the MSSM Higgs
bosons is not quite as advanced as in the SM. Indeed, a full
computation of the two-loop quark-squark-gluino contributions, valid
for arbitrary values of all the relevant particle masses, has not been
made publicly available so far. Moreover, the complexity of such a
computation is going to be reflected in results that will probably be
too lengthy and computer-time-consuming to be efficiently implemented
in event generators. An alternative approach consists in deriving
approximate analytic results that can be easily implemented in
computer codes, and that are valid in specific regions of the MSSM
parameter space such as, e.g., when the Higgs bosons are somewhat
lighter than the squarks and the gluino.

In this paper we presented a new calculation of the two-loop
top-stop-gluino contributions to the form factors for Higgs scalar
production in gluon fusion. We exploited techniques developed in our
earlier computations of the production cross section for the MSSM
Higgs bosons~\cite{DS2,DDVS} to obtain explicit and compact analytic
results based on an asymptotic expansion in the heavy particle masses,
up to and including terms of ${\cal O}(m_\phi^2/\MS^2)$, ${\cal
  O}(\mz^2/\MS^2)$ and ${\cal O}(\mt^2/\MS^2)$.  We compared our new
results with the VHML results of ref.~\cite{DS1}, as well as with the
results of a Taylor expansion in the Higgs mass, up to and including
terms of ${\cal O}(m_\phi^2/\mt^2)$ and ${\cal O}(m_\phi^2/\MS^2)$,
and we discussed the regions of applicability of the different
expansions. We also discussed the effect of choosing different
renormalization schemes for the parameters in the top/stop sector.

From the example presented in section \ref{sec:num} it appears that,
in the case of the heaviest Higgs boson $H$, the use of our asymptotic
expansion becomes mandatory when $\mH$ approaches (or crosses) the
threshold for the production of a real top-quark pair in the loops. It
also appears that choosing the OS scheme for the parameters in the
top/stop sector leads to a milder behavior of the two-loop form factor
$\HHd$ around the threshold. In the case of the lightest Higgs boson
$h$, whose mass is bounded from above in the MSSM, we need not worry
about thresholds. However, our discussion showed that in the $\drbar$
scheme the VHML provides a worse approximation to $\Hhd$ than it does
in the OS scheme.

Finally, we remark that the results derived in this paper for the
production cross section can be straightforwardly adapted to the NLO
computation of the gluonic and photonic decay widths of the MSSM Higgs
bosons, in analogy to what described in section 5 of ref.~\cite{DS1}.


\section*{Acknowledgments}
This work was partially supported by the Research Executive Agency
(REA) of the European Union under the Grant Agreement number
PITN-GA-2010-264564 (LHCPhenoNet).
 


\end{document}